\documentclass{AUJarticle}
\usepackage[cmex10]{amsmath}
\usepackage{amssymb}
\usepackage[utf8x]{inputenc}
\usepackage[nocompress]{cite}
\usepackage{graphicx, multirow, booktabs, color, listings, array}
\usepackage{balance}
\usepackage{xcolor}
\usepackage[export]{adjustbox}
\usepackage{hyperref}

\graphicspath{{figures/}}

\lstdefinelanguage{json}{
  basicstyle=\ttfamily\scriptsize,
  showstringspaces=false,
  breaklines=true,
  frame=single,
  literate=
   *{0}{{{\color{blue}0}}}{1}
    {1}{{{\color{blue}1}}}{1}
    {2}{{{\color{blue}2}}}{1}
    {3}{{{\color{blue}3}}}{1}
    {4}{{{\color{blue}4}}}{1}
    {5}{{{\color{blue}5}}}{1}
    {6}{{{\color{blue}6}}}{1}
    {7}{{{\color{blue}7}}}{1}
    {8}{{{\color{blue}8}}}{1}
    {9}{{{\color{blue}9}}}{1}
    {:}{{{\color{red}{:}}}}{1}
    {,}{{{\color{red}{,}}}}{1}
    {\{}{{{\color{darkgray}{\{}}}}{1}
    {\}}{{{\color{darkgray}{\}}}}}{1}
    {[}{{{\color{darkgray}{[}}}}{1}
    {]}{{{\color{darkgray}{]}}}}{1},
}

\definecolor{darkgray}{rgb}{0.3,0.3,0.3}

\begin{document}

\title{Toward Model-Driven Digital Twin Configuration: Separating Structure, Semantics, and Runtime with SysML+SAREF+Ditto\thanks{Author preprint. Presented at the Workshop on Digital Twin Experiences (DTE), 30th Ada-Europe International Conference on Reliable Software Technologies (AEiC 2026), V\"aster\aa s, Sweden, June 2026; to appear in the \emph{Ada User Journal}.}}

\addauthor{Andrey Sadovykh}
{Softeam, Paris, France}
{andrey.sadovykh@softeam.fr}

\addauthor{Matthew Rusakov}
{Innopolis University, Russia}
{m.rusakov@innopolis.university}

\addauthor{Kirill Korikov}
{Innopolis University, Russia}
{k.korikov@innopolis.university}

\issuev{45}
\issuen{1}
\issued{June 2026}

\shortauthor{A. Sadovykh, M. Rusakov, K. Korikov}
\shorttitle{Toward Model-Driven DT Configuration with SysML+SAREF+Ditto}

\thispagestyle{plain}

\maketitle

\begin{abstract}
Energy communities, neighbourhoods that jointly produce, store, and consume electricity, increasingly rely on \emph{digital twins}: live software copies of physical devices such as a solar panel, a battery, or a heater, kept in sync for monitoring and control. Setting up such a twin is today mostly manual: an engineer hand-writes a configuration whose field names are arbitrary, whose units are implicit, and which carries no machine-readable description of how devices relate, so other systems cannot reliably interpret it. This position paper argues that the task should be split across three complementary tools, each doing what it does best: a modelling language for the system's \emph{structure} (SysML), a standard energy vocabulary for its \emph{meaning} (SAREF4ENER), and a digital-twin platform for the \emph{running representation} (Eclipse Ditto, via the W3C Web-of-Things Thing Description). We give conceptual rules (not yet an executable generator) for deriving a Ditto configuration from a SysML model annotated with that vocabulary, illustrate them on a greenhouse energy-community node modelled in Modelio, and argue that the split improves traceability and configuration consistency, and establishes the semantic hooks needed for interoperability, over hand-written alternatives. We scope the contribution to design-time configuration and discuss its limits.

\vspace{0.5em}
\end{abstract}

\section{Introduction}

Picture a cluster of greenhouses that share a rooftop-solar installation. Each greenhouse has sensors (temperature, humidity) and controllable equipment (heating, ventilation, irrigation), and together they form a small \emph{energy community} that balances local generation and demand. To monitor and coordinate them, each device is given a \emph{Digital Twin} (DT), a live software copy kept in sync with the real hardware, hosted on a DT platform; here we use the open-source \emph{Eclipse Ditto}~\cite{kherbache2022ditto}. Before any of this works, someone must \emph{configure} the platform: tell it what devices exist, what data each produces, and how they relate. Today that configuration is written by hand.

That JSON file carries no shared meaning: attribute names are arbitrary, units implicit, and there is no machine-readable hint that, say, a heater controls a measured temperature. Other systems, an external energy-management system or a sibling greenhouse, therefore cannot reliably interpret it. Scaling to dozens of device types amplifies the problem: without a principled engineering model behind it, each new device type becomes a custom integration effort.

We argue that this task should be split across three complementary tools, each doing what it does best. \emph{SysML}~\cite{omg_sysml}, a standard engineering modelling language, captures the system's \emph{structure}: which devices exist and how they connect. \emph{SAREF4ENER}~\cite{gyrard2021saref}, an ETSI standard \emph{ontology} (a shared, machine-readable vocabulary) for smart-energy devices, supplies interoperable domain \emph{meaning}, agreed terms such as ``power generator'', ``load'', and ``kW''. Ditto, through the W3C Web-of-Things Thing Description (WoT~TD)~\cite{w3c_wot}, provides the deployable \emph{running representation}. From a SysML model annotated with that vocabulary, a small set of rules then derives the Ditto configuration, replacing the opaque hand-written JSON with one carrying explicit types, units, and device relationships.

That this gap is real, and not yet addressed, is borne out by a recent mapping study of model-driven engineering (MDE) for digital twins~\cite{lehner2025mde}: manufacturing and transportation account for over 50\% of applications, energy appears in only 2 of 11 domains, and no surveyed paper targets Eclipse Ditto as a transformation output. This paper focuses on the design-to-configuration direction, at the level of modelling concepts and transformation rules rather than an executable generator, using the greenhouse energy-community node as a running example.

\smallskip\noindent\textbf{Contributions:} (1) a conceptual set of transformation rules from SysML+SAREF4ENER to Ditto/WoT~TD for \emph{initial} twin configuration, the design-to-runtime direction (Table~\ref{tab:rules}); (2) a worked greenhouse example carrying a control relation (\texttt{observes}/\texttt{actsUpon}) end-to-end into the configuration; (3) the position that energy-domain semantics should come from a \emph{standard ontology} (SAREF4ENER) rather than from the modelling language, even under SysML~v2.

\section{Background}

\subsection{SysML and Digital Twin Engineering}

The Systems Modeling Language (SysML)~\cite{omg_sysml} extends the Unified Modeling Language (UML) with blocks, ports, value properties, parametrics, and requirement traceability, making it well-suited as a structural backbone for DT engineering. Wilking et al.~\cite{wilking2022sysml4dt} demonstrate SysML-based DT composition for manufacturing; Pessoa et al.~\cite{pessoa2023mbse} apply Model-Based Systems Engineering (MBSE) to industrial cyber-physical systems (CPS); Ferko et al.~\cite{ferko2025aas} map SysML~v2 to the Asset Administration Shell (AAS), an industry standard for digital asset descriptions. A recent mapping study by Lehner et al.~\cite{lehner2025mde} covering 66 MDE-for-DT papers confirms SysML as a prominent notation.

We use \emph{Modelio}~\cite{modelio}, an open-source MBSE workbench supporting SysML, custom profiles, and API-based model access~\cite{sadovykh2016sysml}, and add to it a SAREF4ENER reference package and conceptual transformation rules toward Ditto configuration.

\subsection{Eclipse Ditto and WoT Thing Description}

Eclipse Ditto~\cite{kherbache2022ditto} is a cloud-native Digital Twin framework that models each physical device as a \emph{Thing} holding \texttt{attributes} (static metadata) and \texttt{features} (live sensor/actuator state, each with named \texttt{properties}). Without extra annotation these are plain key-value pairs: Ditto provides what Crnogorac et al.\ call ``Level~0 domain semantics'', and Pfeiffer et al.\ find that all five DT platforms they evaluate lack built-in semantic typing~\cite{crnogorac2025maturity,pfeiffer2024dt_platforms}. Ditto~3.0+ can attach a \emph{Thing Description}, a standard JSON description of a device and its data~\cite{w3c_wot}, written in JSON-LD: ordinary JSON extended with two keys, \texttt{@type} (tags a value with what kind of thing it is, e.g.\ a power reading) and \texttt{@context} (says where those tags are defined). These are the semantic hooks our approach fills.

\subsection{SAREF and SAREF4ENER}

SAREF~\cite{daniele2016interoperability} is an ETSI standard ontology for IoT interoperability. Its core defines a device taxonomy (\texttt{saref:Device} specializing into \texttt{Sensor}, \texttt{Actuator}, \texttt{Meter}, and \texttt{Appliance}) and a vocabulary of measurement properties (\texttt{saref:Power}, \texttt{saref:Energy}, \texttt{saref:Temperature}, \texttt{saref:Humidity}, and others). The energy extension SAREF4ENER~\cite{gyrard2021saref} adds \texttt{s4ener:Device} (with subclass \texttt{Storage}) and a family of flexibility profiles for demand-response interaction.

\begin{figure}[t]
  \centering
  \includegraphics[width=0.95\columnwidth]{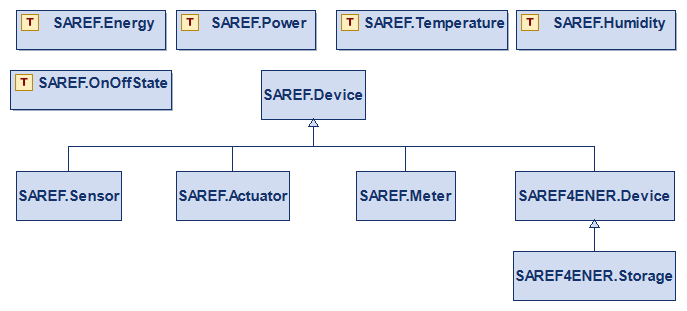}
  \caption{SAREF and SAREF4ENER type hierarchy (Modelio): the DataTypes (top, marked \texttt{T}) serve as value-property types in the greenhouse SysML blocks, while the device taxonomy below provides their block supertypes. A hollow triangle denotes ``is a kind of''.}
  \label{fig:saref}
\end{figure}

Figure~\ref{fig:saref} shows the core of the SAREF/SAREF4ENER type hierarchy. Measurement DataTypes at the top are the types assigned to SysML value properties; the device taxonomy below provides block supertypes (energy-device categories such as \texttt{s4ener:PowerGenerator} and \texttt{s4ener:LoadDevice} are further SAREF4ENER subclasses, used in the example). The \texttt{saref:observes} and \texttt{saref:actsUpon} properties encode sensor-actuator relationships machine-readably.

\subsection{Comparison of Approaches}

Table~\ref{tab:comparison} compares our approach with four alternatives using criteria derived from the needs repeatedly highlighted in the literature: explicit structural modelling, domain semantics, platform-oriented configuration, and a basis for relating design models to runtime twins.

\begin{table}[t]
\centering
\caption{Device description approaches for Energy Community Digital Twins.}
\label{tab:comparison}
\small
\begin{tabular}{lccccl}
\toprule
\textbf{Approach} & \textbf{Struct.} & \textbf{E.\,Sem.} & \textbf{DT\,Cfg.} & \textbf{Trace.} & \textbf{Std.} \\
\midrule
\textbf{Ours}  & \checkmark  & \checkmark  & \checkmark  & \checkmark  & OMG/ETSI \\
WoT-TD         & $\times$    & $\times$    & \checkmark  & $\times$    & W3C      \\
Vorto          & \checkmark  & $\times$    & \checkmark  & $\times$    & Eclipse  \\
IEC~AAS        & \checkmark  & $\times$    & $\times$    & partial     & IEC      \\
NGSI-LD        & $\times$    & partial     & $\times$    & $\times$    & ETSI     \\
\bottomrule
\end{tabular}

\smallskip
{\scriptsize Struct.\,=\,structural design model; E.\,Sem.\,=\,energy domain semantics;
DT\,Cfg.\,=\,Ditto configuration generation; Trace.\,=\,model-runtime traceability;
Std.\,=\,maintaining standards body; NGSI-LD\,=\,ETSI context-information standard. \checkmark\,full, partial, $\times$\,absent;
for our approach \checkmark\ is by construction (conceptual rules), not an executable tool.}
\end{table}

WoT-TD annotates Things but models no design structure; Vorto, an Eclipse toolkit for IoT device models, generates artifacts but carries no energy vocabulary. AAS supports structural modelling, but the SysML-to-AAS mapping of Ferko et al.~\cite{ferko2025aas} keeps only \emph{partial} model-to-runtime traceability; NGSI-LD covers energy concepts only \emph{partially} and offers no engineering structure~\cite{ngsi_ld}. No single notation covers all four criteria, which is why we divide the roles across SysML, SAREF4ENER, and Ditto/WoT~TD.

\section{Approach: SysML+SAREF to Ditto}

SAREF4ENER is imported as a \emph{reference package} (a read-only library of reusable types) in Modelio, so device blocks are typed directly, by specializing SAREF/SAREF4ENER classes and by typing value properties with SAREF measurement types, without heavyweight stereotypes. The \texttt{saref:observes} and \texttt{saref:actsUpon} associations additionally encode sensor-actuator relationships, which the transformation preserves in the target Ditto representation. The next section shows these choices on the greenhouse model.

Table~\ref{tab:rules} maps SysML+SAREF model elements to Ditto/WoT~TD concepts at a conceptual level; formalising them as model-to-text templates (e.g., Acceleo) is future work.

\begin{table}[t]
\centering
\caption{Conceptual transformation rules from SysML+SAREF to Ditto/WoT~TD.}
\label{tab:rules}
\footnotesize
\begin{tabular}{p{2.3cm}p{2.9cm}p{2.0cm}}
\toprule
\textbf{SysML} & \textbf{Ditto/WoT~TD} & \textbf{Purpose} \\
\midrule
Block with SAREF supertype & Feature-level \texttt{@type} & Device category \\
Block name & Thing ID / feature name & Stable naming \\
Value property & \texttt{features.*.properties} & Twin state \\
Property type (\texttt{saref:Power}) & Property-level \texttt{@type} & Measurement meaning \\
Unit annotation & \texttt{unit} field & Explicit units \\
\texttt{saref:observes} & Semantic link & Sensor-to-property \\
\texttt{saref:actsUpon} & Desired property & Actuator intent \\
Block multiplicity & Repeated features / Things & Many instances \\
\bottomrule
\end{tabular}
\end{table}

The transformation preserves semantics by carrying the SAREF URI into \texttt{@type}, so model and twin share one ontological identity and every Ditto property traces back to a named SysML attribute; the WoT~TD \texttt{@context} makes the result self-describing and JSON-LD-processable. At this stage we generate the configuration text correctly but do not yet check it with automated reasoning or validation tools (such as OWL or SHACL), nor verify model-to-twin round-trips; that is future work.

\section{Greenhouse Example}

\subsection{Scenario}

We illustrate the approach on a greenhouse serving as a single Energy Community node with four device types: \texttt{TemperatureSensor}, \texttt{HVACSystem}, \texttt{SolarPanel}, and \texttt{GridConnection}. Without the proposed approach, each device is configured by hand-editing Ditto JSON, with attribute names and units chosen ad hoc per device; we instead derive that configuration from a single SysML model.

\subsection{SysML+SAREF Model}

\begin{figure*}[t]
  \centering
  \includegraphics[width=0.98\textwidth]{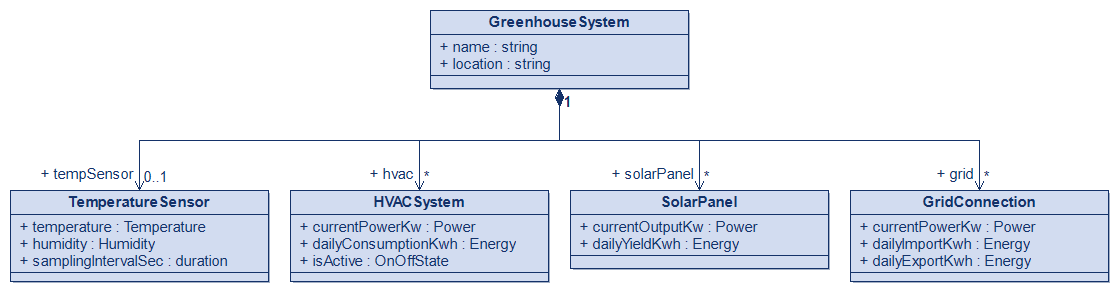}
  \caption{SysML BDD of the greenhouse Energy Community node, modelled in Modelio. \texttt{GreenhouseSystem} decomposes into four device blocks whose value properties are typed with SAREF measurement types; the diamond denotes composition and \texttt{*}/\texttt{0..1} are multiplicities. The diagram shows the full block detail; the tables and listings use a mapped subset, and the SAREF supertypes and \texttt{observes}/\texttt{actsUpon} relations live in the model but are not drawn here.}
  \label{fig:bdd}
\end{figure*}

Figure~\ref{fig:bdd} shows the greenhouse Block Definition Diagram (BDD). The \texttt{GreenhouseSystem} block at the top owns four device blocks, drawn left to right as \texttt{TemperatureSensor}, \texttt{HVACSystem}, \texttt{SolarPanel}, and \texttt{GridConnection}. Each device block specializes a SAREF/SAREF4ENER supertype (via a generalization to the reference package) and lists value properties typed with SAREF measurement types: for example, \texttt{SolarPanel} carries \texttt{currentOutputKw\,::\,Power} and \texttt{dailyYieldKwh\,::\,Energy}, while \texttt{GridConnection} carries \texttt{dailyImportKwh} and \texttt{dailyExportKwh}, both typed \texttt{Energy}. In the same model \texttt{TemperatureSensor} \texttt{observes} \texttt{saref:Temperature} and \texttt{HVACSystem} \texttt{actsUpon} it, capturing the control relationship in the model. Table~\ref{tab:greenhouse} summarises the per-block mapping obtained by applying the rules of Table~\ref{tab:rules}: the container block becomes the Thing and each device block a feature.

\begin{table}[t]
\centering
\caption{Mapping decisions for the greenhouse blocks.}
\label{tab:greenhouse}
\scriptsize
\begin{tabular}{@{}p{1.5cm}p{1.0cm}p{2.0cm}p{2.0cm}@{}}
\toprule
\textbf{SysML block} & \textbf{Ditto} & \textbf{SAREF \texttt{@type}} & \textbf{Rationale} \\
\midrule
\texttt{Greenhouse\-System} & Thing & n/a & Container is the Thing \\
\texttt{SolarPanel} & feat. \texttt{solar} & \texttt{s4ener:Power\-Generator} & Generation; power/energy \\
\texttt{HVAC\-System} & feat. \texttt{hvac} & \texttt{s4ener:Load\-Device} & Load; \texttt{actsUpon} Temp. \\
\texttt{Temperature\-Sensor} & feat. \texttt{temp} & \texttt{saref:Sensor} & Sensing; \texttt{observes} Temp. \\
\texttt{Grid\-Connection} & feat. \texttt{grid} & \texttt{saref:Meter} & Bidirectional metering \\
\bottomrule
\end{tabular}

\smallskip
{\scriptsize feat.\,=\,Ditto feature; n/a\,=\,no device-category type (the Thing itself).}
\end{table}

\subsection{Generated Ditto Configuration}

Listing~\ref{lst:json} applies the rules of Table~\ref{tab:rules} to the \texttt{SolarPanel} block of Figure~\ref{fig:bdd}. Without the approach (top), the feature holds two opaque numbers whose names, units, and device role are left to convention. With the approach (bottom), the derived feature carries the device category (\texttt{s4ener:PowerGenerator}), each property's measurement meaning (\texttt{saref:Power}, \texttt{saref:Energy}), and an explicit \texttt{unit}, with names matching the SysML attributes. The benefit is direct: an external energy-management system can interpret the twin without prior knowledge of this particular configuration. A JSON-LD \texttt{@context} (omitted from the listings for brevity) binds the \texttt{saref:} and \texttt{s4ener:} prefixes to their ETSI ontology URIs, so the \texttt{@type} values resolve to globally shared definitions rather than local strings.

\begin{lstlisting}[language=json,float=tbp,caption={Solar-panel feature before (top) and after (bottom) applying the transformation rules.},label={lst:json}]
{ "features": { "solar": {
  "properties": { "power": 2.4, "energy": 12.1 }
}}}
//  ===== derived (after applying the rules) =====
{ "features": { "solar": {
  "@type": "s4ener:PowerGenerator",
  "properties": {
    "currentOutputKw": {
      "@type": "saref:Power", "value": 2.4, "unit": "kW" },
    "dailyYieldKwh": {
      "@type": "saref:Energy", "value": 12.1, "unit": "kWh" }
}}}}
\end{lstlisting}

Listing~\ref{lst:hvac} shows the actuator side, derived from the \texttt{HVACSystem} block. Its measured state goes to \texttt{properties}, while the \texttt{actsUpon} relation becomes a \texttt{desiredProperties} entry naming the controlled quantity (\texttt{saref:Temperature}). The twin thus records not only what the device reports but what it is meant to control, making the sensor-to-actuator loop (\texttt{TemperatureSensor}\,$\rightarrow$\,\texttt{HVACSystem}) explicit in the configuration.

\begin{lstlisting}[language=json,float=tbp,caption={HVAC feature: measured state under \texttt{properties} and the \texttt{actsUpon} target under \texttt{desiredProperties}.},label={lst:hvac}]
{ "features": { "hvac": {
  "@type": "s4ener:LoadDevice",
  "properties": {
    "currentPowerKw": {
      "@type": "saref:Power", "value": 1.8, "unit": "kW" },
    "dailyConsumptionKwh": {
      "@type": "saref:Energy", "value": 9.4, "unit": "kWh" },
    "isActive": {
      "@type": "saref:OnOffState", "value": true } },
  "desiredProperties": {
    "setTemperature": {
      "@type": "saref:Temperature", "value": 22.0, "unit": "degC" } }
}}}
\end{lstlisting}

\section{Discussion}

DT engineering spans structural, semantic, and runtime concerns that no single notation handles well, which is why we assign each to the tool suited to it. Bordeleau et al.~\cite{bordeleau2020mde} identify model heterogeneity, bidirectional synchronization, and model evolution as three central challenges for MDE-based DT engineering, all of which call for coordination across notations rather than one monolithic model. Our split responds to those challenges, and to three further gaps in the literature: architectural fragmentation in DT engineering, lack of native semantics in IoT twin platforms, and underrepresentation of the energy domain in existing SysML-to-platform mappings. As Listing~\ref{lst:json} shows, it turns opaque hand-written JSON into a configuration with explicit device categories, measurement semantics, and units, with sensor-actuator intent (\texttt{TemperatureSensor}$\rightarrow$\texttt{HVACSystem} via \texttt{saref:Temperature}) encoded machine-readably.

\smallskip\noindent\textbf{Why SAREF, given SysML~v2?} One could define an energy vocabulary directly as a SysML~v2~\cite{ferko2025aas} model library and skip the ontology. The difference is that such a library is \emph{local} to the project, whereas SAREF4ENER is an ETSI standard that external energy-management systems already consume; the question is therefore not v1-versus-v2 but local-versus-standardised semantics. SysML~v2's formal, machine-readable models sharpen interchange between engineering tools, but supply no agreed energy terms such as \texttt{PowerGenerator}, \texttt{Power}, or \texttt{kWh}. SAREF4ENER supplies that shared meaning, so the two are complementary: SysML (v1 or v2) contributes the structural layer and SAREF4ENER the standardised domain semantics, and our rules treat the SAREF reference package as the semantic source, agnostic to the SysML version.

\smallskip\noindent\textbf{Scalability and reuse.} The SAREF reference package is imported once and reused across device types, and the block-multiplicity rule of Table~\ref{tab:rules} is designed to derive many twin instances from one typed block, so the same model that describes a single greenhouse should extend to a community of nodes by instantiation rather than hand-editing JSON per device. The split is also intended to be domain-portable: the rules stay fixed and only the reference package changes, in principle swapping SAREF4ENER for another ETSI SAREF extension (S4BLDG for buildings, S4AGRI for farming, S4CITY for urban systems)~\cite{daniele2016interoperability}, keeping the energy specifics in the ontology rather than the rules. Demonstrating both at scale is future work.

The approach introduces modelling overhead and assumes familiarity with SysML and ontology concepts; some choices, notably the partitioning of blocks into Things versus features in larger systems, remain design decisions rather than universal rules, and SAREF4ENER does not cover every energy-community concept. Even so, the split provides a clearer architectural basis than starting from runtime JSON, and extends naturally to flexibility profiles via the \texttt{FlexibilityProfile} and \texttt{PowerSequence} concepts.

\section{Related Work}

Four strands of work bear on this paper. SysML-based DT engineering is effective for structural design and lifecycle traceability, but usually targets manufacturing or platforms other than Ditto~\cite{wilking2022sysml4dt,ferko2025aas,sadovykh2016sysml}. DT-platform research shows Eclipse Ditto is practical at runtime, but its native model is intentionally lightweight and provides no energy-domain semantics~\cite{kherbache2022ditto,pfeiffer2024dt_platforms}. SAREF and SAREF4ENER offer the standard vocabulary for smart-energy assets, yet ontology-centred work rarely ties those concepts back to engineering models~\cite{daniele2016interoperability}. The closest precedent, Heithoff et al.~\cite{heithoff2025digital_shadow}, stores SysML element names in Ditto metadata for traceability, but without energy semantics or SAREF integration. No prior work describes this combination for energy-community digital twins.

\section{Conclusion}

Separating structure (SysML), semantics (SAREF4ENER), and runtime representation (Ditto/WoT~TD) yields configuration artifacts that are traceable and semantically grounded, without requiring any single notation to exceed its design scope. The transformation rules have concrete consequences: explicit types, units, and sensor-actuator relations that would otherwise stay implicit in hand-written JSON.

Next steps: formalising the rules as executable model-to-text templates, validating them across more SAREF4ENER device types and through interoperability experiments, extending to demand-response flexibility profiles, and supporting runtime reflection from Ditto back to the SysML model.

\section*{Acknowledgment}
This work is partly supported by MATISSE (Chips Joint Undertaking, grant No.\ 101140216).

\balance

\end{document}